\documentclass[aps,prb,twocolumn,nofootinbib,notitlepage,superscriptaddress,longbibliography,preprintnumbers]{revtex4-2}
\usepackage{times}
\usepackage{graphicx}
\usepackage[whole]{bxcjkjatype} % for Japanese
\usepackage[T1]{fontenc}

\usepackage[top=30truemm,bottom=30truemm,left=25truemm,right=25truemm]{geometry}

\usepackage[colorlinks=true,linktoc=page,citecolor=red,linkcolor=blue]{hyperref}	
\graphicspath{{./figs/}}  % pass for figs
\usepackage[usenames]{xcolor}
\usepackage{amsmath,amssymb,amsfonts}	
\usepackage{bm,braket,array}
\usepackage{multirow}
\usepackage[all]{xy}
\usepackage{amsthm}
\usepackage{amscd}
\usepackage{slashed} %for D slash
\usepackage{calligra} %for symbols of rod groups
\usepackage[normalem]{ulem} % for edit
\theoremstyle{definition}

\theoremstyle{remark}

\def\sgn{{\rm sgn\,}}

\newcommand{\g}{\gamma}

\newcommand{\bk}{{\bm{k}}}

\def\widebar{\accentset{{\cc@style\underline{\mskip10mu}}}} %widebar
\def\wideubar{\underaccent{{\cc@style\underline{\mskip10mu}}}} %wideubar

\begin{document}
\title{A discrete formulation for three-dimensional winding number}
\author{Ken Shiozaki}
\affiliation{Center for Gravitational Physics and Quantum Information, Yukawa Institute for Theoretical Physics, Kyoto University, Kyoto 606-8502, Japan}
\date{\today}
\preprint{YITP-24-14}

\begin{abstract}
For a smooth map $g: X \to U(N)$, where $X$ is a three-dimensional, oriented, and closed manifold, the winding number is defined as $W_3 = \frac{1}{24\pi^2} \int_{X} \mathrm{Tr}\left[(g^{-1}dg)^3\right]$. We present a discrete formulation to compute $W_3$ based on the concept of $\theta$-gaps. Our approach provides a robust scheme that is directly applicable even to systems with accidental or symmetry-enforced degeneracies. Furthermore, we define two versions of the discrete flux: a simple unmodified flux that is highly practical and almost always quantized for fine grids, and a modified flux that strictly ensures integer quantization.
\end{abstract}

\maketitle
%\tableofcontents
\parskip=\baselineskip

{\it Introduction---}
Consider a three-dimensional closed and oriented manifold $X$. 
For a smooth map $g: X \to U(N)$, with $U(N)$ representing the group of $N \times N$ unitary matrices, the winding number, an integer value, is defined by the following expression:
\begin{align}
W_3[g] &= \frac{1}{2\pi} \int_{X} H \in \mathbb{Z}, \\
H &= \frac{1}{12\pi} \mathrm{Tr}\left[(g^{-1}\mathrm{d}g)^3\right]. 
\end{align}
The winding number $W_3[g]$ is pivotal in various branches of physics, including topological band theory, where it acts as the topological invariant for three-dimensional superconductors with time-reversal symmetry~\cite{volovik2003universe, Schnyder=Ryu=Furusaki=Ludwig08}, and in non-Abelian (lattice) gauge theory, appears in instanton number calculations~\cite{nakahara2018geometry,luscher1982topology}. 
Often in these applications, the function $g$ is defined only on a finite set of lattice points for numerical analysis. 
Therefore, an efficient numerical formulation with lattice approximation of manifolds is an important issue.

For the first Chern number $ch_1$ of a line bundle with connection, a sort of two-dimensional counterpart of the winding number, a well-established discrete formulation with manifest quantization exists~\cite{Fujiwara=Suzuki=Wu01,FHS05}. 
Furthermore, discrete line bundles over finite simplicial complexes have been explored, especially concerning applications in computer graphics~\cite{Knoppel2016}. 
For the three-dimensional winding number, a discrete formulation that ensures integer quantization was pioneered by H\"{o}ckendorf, Alvermann, and Fehske~\cite{Hockendorf_2017}. Their algorithm relies on tracking individual bands by matching eigenvalues and eigenvectors between adjacent discretization points. While effective, this matching process can become intricate when the unitary map $g$ features degenerate eigenstates, whether accidental or symmetry-enforced. 

This paper develops an alternative but related discrete method for evaluating $W_3[g]$ based on the concept of a $\theta$-gap. By bypassing the need to track individual bands across adjacent points, our formulation remains robust and directly applicable even in the presence of degeneracies. 
We introduce two versions of the discrete flux: an unmodified flux $\Phi_p$ that is determined solely by the local information on each plaquette, and a modified flux $\tilde{\Phi}_p$ that strictly ensures integer quantization by incorporating the $\theta$-gap information from cubes surrounding the links to rearrange the discrete Berry connection. 
While $\tilde{\Phi}_p$ provides a rigorous topological invariant, we demonstrate that the simpler $\Phi_p$ is almost always quantized for sufficiently fine grids, making it a highly practical choice for numerical evaluations.

\begin{figure*}[!]
\centering
\includegraphics[width=0.8\linewidth, trim=0cm 0cm 0cm 0cm]{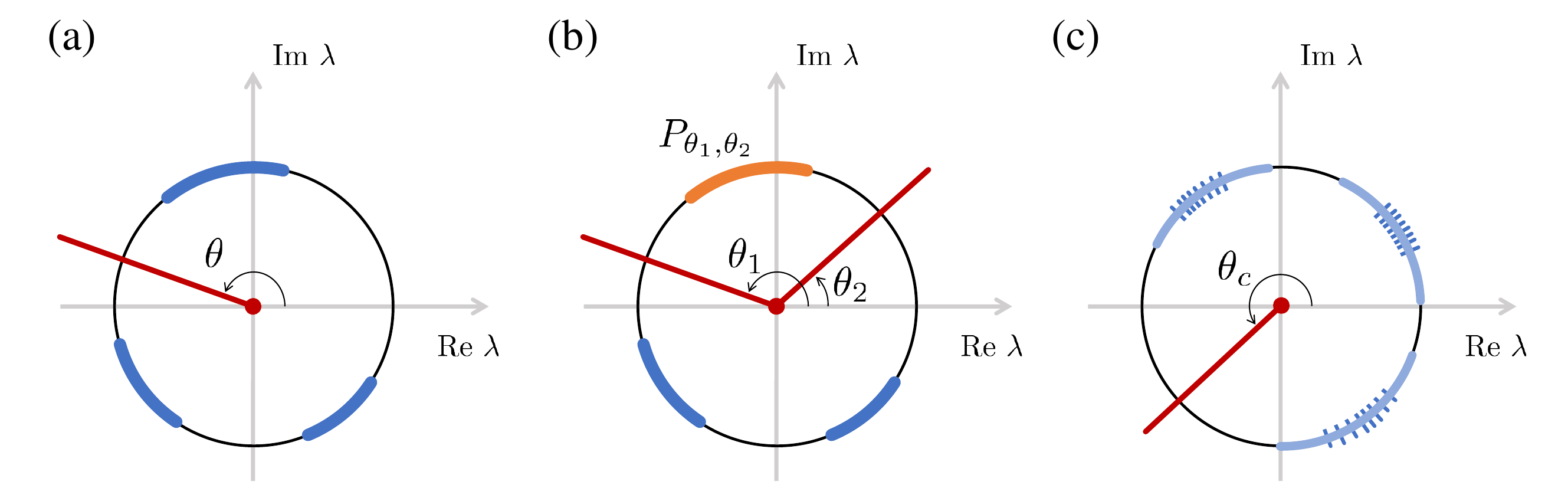}
\caption{(a) Illustration of a $\theta$-gap. (b) Projection onto the eigenspace between two $\theta$-gaps. (c) Smearing eigenvalues over a finite set of vertices. 
In each panel, the unit circle in the complex plane is depicted, with blue arcs representing the spectrum of the $U(N)$-valued matrix $g(x)$ within a local region. Red lines mark the locations of $\theta$-gaps.}
\label{fig:gap}
\end{figure*}

{\it Formulation---}
Given that $H$ is a closed three-form, it is locally exact, meaning that for a local patch, there exists a two-form $B$ such that $H = dB$. To construct $B$ explicitly, we introduce a gap condition for elements of $U(N)$. A matrix $g \in U(N)$ exhibits a $\theta$-gap if none of its eigenvalues are $e^{i\theta}$ for a given real number $\theta \in [0,2\pi)$~\cite{gawedzki2015bundlegerbestopologicalinsulators,CDFGT15}, as illustrated in Fig.~\ref{fig:gap} (a). Furthermore, we define $\log_\theta z$ for  nonzero complex number $z \in \mathbb{C}^\times$ as
\begin{align}
    \log_\theta z = \log |z| + i \arg z,\quad \theta \leq \arg z < \theta + 2\pi. 
\end{align}
For two distinct $\theta$-gaps, $\theta_1$ and $\theta_2$, the following relation holds:
\begin{widetext}
\begin{align}
    \log_{\theta_1} z - \log_{\theta_2} z 
    &= \begin{cases}
    2\pi i \times \mathrm{sgn}(\theta_1-\theta_2) & (\min(\theta_1,\theta_2) < \arg z < \max(\theta_1,\theta_2)), \\
    0 & (\mathrm{otherwise}). 
    \end{cases}
    \label{eq:log_diff}
\end{align}
\end{widetext}
Consider $U \subset X$ as a three-dimensional subspace where $g(x)$ maintains a $\theta$-gap for $x \in U$. Let $\g(x) = (u_1(x),\dots,u_N(x)) \in U(N)$ be a unitary matrix diagonalizing $g(x)$, i.e., $g(x) = \g(x) \Lambda(x) \g(x)^\dagger$ with $\Lambda(x) = \mathrm{diag}(\lambda_1,\dots, \lambda_N)$, where $\lambda_n \in U(1)$ for $n=1,\dots,N$. The exact form $B_\theta$ is given by~\cite{GR02}
\begin{align}
    B_\theta &= Q + R_\theta, \\
    Q &= \frac{1}{4\pi} \mathrm{Tr}[\g^{-1}d\g \Lambda \g^{-1}d\g \Lambda^{-1}], \\
    R_\theta &= \frac{1}{2\pi} \mathrm{Tr}[\log_\theta \Lambda (\g^{-1}d\g)^2].
\end{align}
Note that $Q$ is independent of $\theta$, whereas $R_\theta$ depends on it.
It is evident that $X$ can be covered by $\{U_i\}_i$ such that in each patch $U_i$, $g(x)$ exhibits a specific $\theta$-gap $\theta_i$ for $x \in U_i$.  

The unitary matrix $\gamma$ is not unique because it can be transformed as $\gamma \mapsto \gamma W$, where $W \in U(N)$ commutes with $\Lambda$, i.e., $W\Lambda W^{-1} = \Lambda$. 
However, this ambiguity does not affect $Q$ and $R_\theta$: 
To clarify, consider the $N$ eigenvalues of $\Lambda$ grouped into blocks of $|I|$ degenerate eigenvalues, each associated with a common eigenvalue $\lambda_I$, so that $\Lambda$ can be expressed as $\Lambda = \bigoplus_I \lambda_I \mathbf{1}_{|I|}$.
Introduce a block matrix notation $A_{IJ} = (u_{i}^\dagger du_j)_{i \in I,j \in J}$. 
The transformation matrix $W$ is expressed as $W = \bigoplus_I W_I$ as well, with $W_I \in U(|I|)$, modifying $A_{IJ}$ to $A_{IJ} = W_I^\dagger A_{IJ} W_J + \delta_{IJ} W_I^{-1} d W_I$. 
Consequently, $Q$ and $R_\theta$ can be represented as:
\begin{align}
    Q &= \frac{1}{4\pi} \sum_{I,J; I \neq J} \mathrm{Tr}_I [A_{IJ} A_{JI}] \lambda_J \lambda_I^{-1}, \\
    R_\theta &= \frac{1}{2\pi} \sum_{I,J; I \neq J} \mathrm{Tr}_I [A_{IJ}A_{JI}] \log_\theta \lambda_I,
\end{align}
where $\mathrm{Tr}_I$ denotes the trace over indices $i \in I$. 
In the summation $\sum_{I,J}$, terms with $I = J$ can be excluded due to $\mathrm{Tr}_I[(A_{II})^2] = 0$. 
This demonstrates the invariance of $Q$ and $R_\theta$ under the transformation $\gamma \mapsto \gamma W$.

\begin{figure}[!]
\centering
\includegraphics[width=0.6\linewidth, trim=0cm 0cm 0cm 0cm]{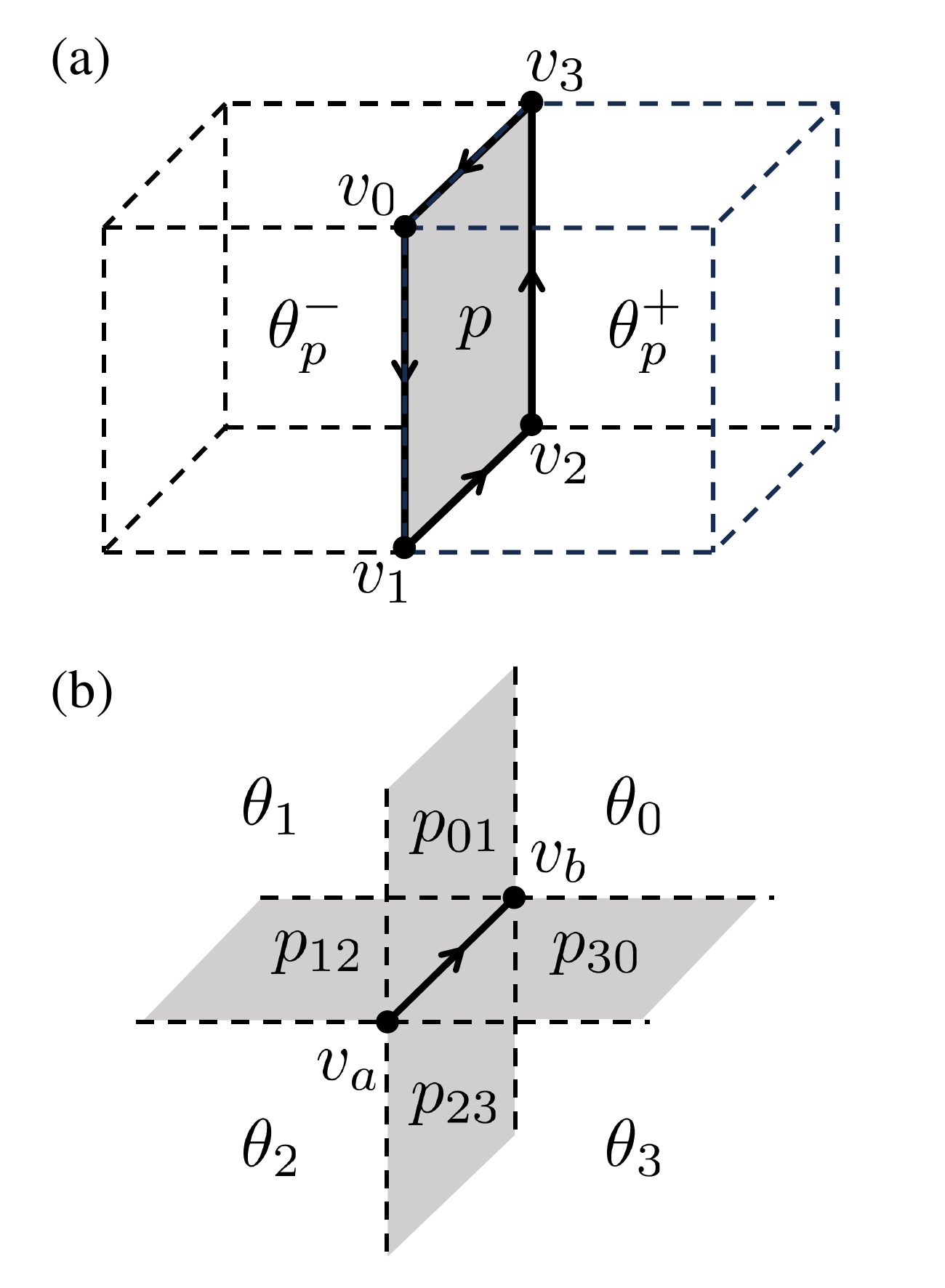}
\caption{
(a) A plaquette $p$ within the cubic lattice, showing $\theta_p^+$ and $\theta_p^-$ as the $\theta$-gaps of cubes adjacent to $p$, aligned parallel and anti-parallel to $p$'s normal vector, respectively. The vertices $v_0, v_1, v_2,$ and $v_3$ are sequentially labeled around the perimeter of plaquette $p$. 
(b) An edge $v_av_b$ of the cubic lattice, illustrating the $\theta$-gaps of cubes adjacent to the edge $v_av_b$. 
}
\label{fig:plaquette}
\end{figure}

Another noteworthy aspect is that the difference in $B_\theta$ between two $\theta$-gaps is a total derivative. 
For $0\leq \theta_1, \theta_2 < 2\pi$, and using (\ref{eq:log_diff}), it follows that:
\begin{align}
    B_{\theta_1} - B_{\theta_2} = d\alpha_{\theta_1,\theta_2},
\end{align}
where 
\begin{align}
    \alpha_{\theta_1,\theta_2} &= - i\ \mathrm{sgn}(\theta_1-\theta_2) \mathrm{Tr} [P_{\theta_1,\theta_2} \gamma^{-1}d\gamma] \nonumber \\
    &= - i\ \mathrm{sgn}(\theta_1-\theta_2) \nonumber \\
    &\quad \times \sum_{n; \min(\theta_1,\theta_2)<\arg \lambda_n < \max(\theta_1,\theta_2)} u_n^\dagger du_n,
\end{align}
with
\begin{align}
    &P_{\theta_1,\theta_2} = \mathrm{diag}(p_1, \dots, p_N), \nonumber \\
    &p_n = \begin{cases} 1 & (\min(\theta_1,\theta_2) < \arg \lambda_n < \max(\theta_1,\theta_2)) \\ 0 & (\mathrm{otherwise}) \end{cases}
\end{align}
the orthogonal projection onto the eigenspace for eigenvalues that fulfill $\min(\theta_1,\theta_2)<\arg \lambda_n < \max(\theta_1,\theta_2)$. 
Note that the projection $P_{\theta_1,\theta_2}$ is $\bk$-independent in the eigenstate basis, 
whereas in the original basis it reads $\gamma P_{\theta_1,\theta_2} \gamma^\dagger=\sum_{n; \min(\theta_1,\theta_2)<\arg \lambda_n < \max(\theta_1,\theta_2)} u_n u_n^\dagger$. 
In cases where $\theta_1=\theta_2$ or no eigenvectors meet the condition $\min(\theta_1,\theta_2)<\arg \lambda_n < \max(\theta_1,\theta_2)$, $\alpha_{\theta_1,\theta_2}$ is simply null.
Now, we express the winding number $W_3[g]$ as a sum of line integrals, utilizing a cubic lattice approximation $L$ of the manifold $X$. (Any simplicial decomposition is equally valid.) 
For each cube $c$ of the lattice $L$, we select $\theta_c \in [0,2\pi)$ such that $g_x$ for $x \in c$ exhibits a $\theta$-gap of $\theta_c$. Thus, $W_3[g]$ can be reformulated as a sum of integrals over all plaquettes:
\begin{align}
    W_3[g] &= \frac{1}{2\pi} \sum_{c} \int_{c} dB_{\theta_c} \nonumber \\
    &= \frac{1}{2\pi} \sum_{c} \int_{\partial c} B_{\theta_c}  \nonumber \\
    &= \frac{1}{2\pi} \sum_p \int_{p} (B_{\theta_p^-}-B_{\theta_p^+}).
    \label{eq:W3_red_1}
\end{align}
Here, $\sum_p$ runs over all plaquettes $p$ in the lattice $L$, with each $p$ being oriented. The gap parameters $\theta_p^+$ and $\theta_p^-$ correspond to the cubes adjacent to plaquette $p$, in directions parallel and antiparallel to $p$'s normal vector, respectively, as depicted in Fig.~\ref{fig:plaquette} (a). 
This expression further simplifies to a sum of line integrals:
\begin{align}
    W_3[g] &= \frac{1}{2\pi} \sum_p \int_{p} d\alpha_{\theta_p^-,\theta_p^+} \nonumber \\
    &= \frac{1}{2\pi} \sum_p \oint_{\partial p} \alpha_{\theta_p^-,\theta_p^+}.
    \label{eq:W3_red_2}
\end{align}
The transition to the last expression is contingent upon $\alpha_{\theta_p^-,\theta_p^+}$ being smoothly defined across plaquette $p$. 
When $\alpha_{\theta_p^-,\theta_p^+}$ is solely determined along the loop $\partial p$ bordering plaquette $p$, a $2\pi$ ambiguity may arise from large gauge transformations $u_n \to u_n e^{i\chi_n}$, where $\oint_{\partial p} d\chi_n = 2\pi$, potentially altering $W_3[g]$ by an integer. 
However, if the cubic lattice $L$ is sufficiently fine relative to the spatial variations of $g$, the loop integral $\oint_{\partial p} \alpha_{\theta_p^-,\theta_p^+}$ will take values close to 0, within the range of $-\pi$ to $\pi$ modulo $2\pi$. 
This allows it to be treated as an $\mathbb{R}$-valued quantity without the $2\pi$ ambiguity.

We claim that the winding number, as expressed in (\ref{eq:W3_red_2}), can be approximately calculated using only the diagonalizing matrices $\gamma(v)$ at the vertices $v$ of the lattice $L$, while ensuring that $W_3[g]$ remains quantized.
Diagonalizing $g(v)$ at vertices $v \in L$ yields the set of pairs of eigenvectors and eigenvalues $\{u_n(v),\lambda_n(v)\}_{n=1,\dots,N}$ for each vertex $v$. 
The eigenvalues \(\lambda_n(v)\) are labeled such that their angles are in ascending order, satisfying 
\begin{align}
    0 \leq \arg(\lambda_1(v)) \leq \cdots \leq \arg(\lambda_N(v)) < 2\pi.
\end{align} 
This labeling prevents the sign factor that arises from reordering the eigenvectors over vertices when the Berry connection is evaluated as the inner product of eigenvectors among two adjacent vertices. 
Note also that eigenvalues $\lambda$ near $0$ may reorder significantly under minor perturbations, yet this does not contribute to the discrete formula below. 
We introduce a notation for the set of eigenstates between two angles: 
For a given pair of gap parameters $0 \leq \theta_1, \theta_2 < 2\pi$, we define 
\begin{align}
    \gamma_{\theta_1,\theta_2}(v)
    := \left( u_{n_1}(v), u_{n_1+1}(v), \dots, u_{n_2}(v) \right),
\end{align}
where the eigenstates satisfy $\min(\theta_1,\theta_2) < \arg \left(\lambda_{n_1}(v)\right) \leq \cdots \leq \arg \left(\lambda_{n_2}(v)\right) < \max(\theta_1,\theta_2)$.

The gap parameter $\theta_c$ for each cube $c$ is numerically determined as follows: 
From the eight corner vertices of cube $c$, denoted as $v \in c$, we consider the set of $8N$ eigenvalues $\{\lambda_n(v)\}_{v \in c,n=1,\dots,N}$. 
By smearing all eigenvalues $\lambda_n(v)$, we have a set of intervals:
\begin{align}
    I_c &= \bigcup_{v \in c,n=1,\dots,N} \Bigg\{ \arg (\lambda_{n}(v) e^{i\delta \phi}) \in [0,2\pi) \Bigg| \nonumber\\
    &\hspace{30pt} -\frac{\beta}{2N} < \delta \phi < \frac{\beta}{2N} \Bigg\}.
\end{align}
Here, $0 < \beta < 1$ is a constant smearing parameter ensuring that near eigenvalues fall within the same smeared interval. 
For example, we can set as $\beta = 1/2$. 
We select a $\theta$ from the complement $[0,2\pi) \backslash I_c$ to serve as $\theta_c$ for cube $c$. 
(See Fig.~\ref{fig:gap} (c) for visualization.) 
Alternatively, one can simply define $\theta_c$ as the midpoint of the largest gap among the $8N$ angles $\{\arg(\lambda_n(v))\}_{v \in c,n=1,\dots,N}$.
With $\theta$-gaps for all cubes in lattice $L$ thus defined, the gap parameters $\theta_p^+$ and $\theta_p^-$ for each plaquette $p$ are specified. 

For the corner vertices $v_0, v_1, v_2, v_3$ of a plaquette $p$, we define the eigenframes as
\begin{align}
    \gamma_{\theta_p^+,\theta_p^-}(v_a) = \left( u_{n_1}(v_a), \dots, u_{n_2}(v_a) \right)
\end{align}
for $a=0,1,2,3$, comprising the $n_2-n_1$ eigenvectors with eigenvalues between $\theta_p^+$ and $\theta_p^-$. The nonnegative integer $n_2-n_1$, representing the count of eigenvalues between $\theta_p^+$ and $\theta_p^-$, should be consistent across the four vertices $v_0, v_1, v_2, v_3$ of plaquette $p$, provided that the lattice $L$ is sufficiently fine. 
The line integral $\oint_{\partial p} \alpha_{\theta_p^-,\theta_p^+}$ can then be approximated by the discrete Berry phase as
\begin{align}
    \oint_{\partial p} \alpha_{\theta_p^-,\theta_p^+}
    &\cong \mathrm{Arg} \Big[e^{iA_{v_3v_0 \to p}} e^{iA_{v_2v_3\to p}} 
    \nonumber\\
    &\times e^{iA_{v_1v_2\to p}} e^{iA_{v_0v_1\to p}} \Big] =: \Phi_p,
    \label{eq:cont_p}
\end{align}
where
\begin{align}
    e^{iA_{v_av_b\to p}} = \det \left[ \gamma_{\theta_p^+,\theta_p^-}(v_b)^\dagger \gamma_{\theta_p^+,\theta_p^-}(v_a) \right]^{\mathrm{sgn}(\theta_p^+-\theta_p^-)}
\end{align}
is the approximate edge $U(1)$ connection that contributes to the flux piercing the plaquette $p$. Here, $\mathrm{Arg}$ denotes the principal value, with $-\pi < \mathrm{Arg}\, z < \pi$.

Although the sum of fluxes 
\begin{align}
    W^{\rm dis}_3[g] := \frac{1}{2\pi} \sum_p \Phi_p
\end{align}
provides a good approximation of the three-dimensional winding number $W_3[g]$, it is not generally quantized. To ensure that the sum is an integer, we express the sum of fluxes $\Phi_p$ modulo $2\pi$ in terms of edge contributions as follows:
\begin{widetext}
\begin{align}
\sum_p \Phi_p
&\equiv \sum_{v_av_b \in \{\mathrm{edges}\}} \mathrm{Arg} \Bigg[
e^{iA_{v_av_b \to p_{01}}} e^{iA_{v_av_b\to p_{12}}} e^{iA_{v_av_b \to p_{23}}} e^{iA_{v_av_b\to p_{30}}}  
\Bigg] \\
&\equiv \sum_{v_av_b \in \{\mathrm{edges}\}} \mathrm{Arg} \Bigg[
\det\left[ \gamma_{\theta_0,\theta_1}(v_b)^\dagger \gamma_{\theta_0,\theta_1}(v_a) \right]^{\mathrm{sgn}(\theta_0-\theta_1)}
\det\left[ \gamma_{\theta_1,\theta_2}(v_b)^\dagger \gamma_{\theta_1,\theta_2}(v_a) \right]^{\mathrm{sgn}(\theta_1-\theta_2)} \nonumber\\
&\hspace{60pt} \times 
\det\left[ \gamma_{\theta_2,\theta_3}(v_b)^\dagger \gamma_{\theta_2,\theta_3}(v_a) \right]^{\mathrm{sgn}(\theta_2-\theta_3)}
\det\left[ \gamma_{\theta_3,\theta_0}(v_b)^\dagger \gamma_{\theta_3,\theta_0}(v_a) \right]^{\mathrm{sgn}(\theta_3-\theta_0)}
\Bigg] \quad \bmod 2\pi. 
\label{eq:edge}
\end{align}
Here, $v_av_b$ denotes an individual edge, $p_{01}, p_{12}, p_{23}, p_{30}$ are the plaquettes adjacent to the edge $v_av_b$, and $\theta_0, \theta_1, \theta_2, \theta_3$ are the gap parameters such that the plaquette $p_{j,j+1}$ is adjacent to the two cubes with gap parameters $\theta_j$ and $\theta_{j+1}$. See Fig.~\ref{fig:plaquette} (b).
An edge contribution ${\rm Arg}\left[ e^{iA_{v_av_b \to p_{01}}} e^{iA_{v_av_b\to p_{12}}} e^{iA_{v_av_b \to p_{23}}} e^{iA_{v_av_b\to p_{30}}}\right]$ may not be zero. This occurs when an eigenframe $\gamma_{p_{j,j+1}}(v_a)$ for a plaquette $p_{j,j+1}$ includes two or more other eigenframes.

For instance, consider the situation where $\theta_0 < \theta_1 < \theta_2 = \theta_3$ and both the number of eigenvalues between $\theta_0$ and $\theta_1$, and $\theta_1$ and $\theta_2$, are nonzero. Then, $\gamma_{\theta_3,\theta_0}(v) = \left(\gamma_{\theta_0,\theta_1}(v),\gamma_{\theta_1,\theta_2}(v) \right)$ holds for $v = v_a, v_b$, and we have 
\begin{align}
e^{iA_{v_av_b \to p_{01}}} e^{iA_{v_av_b\to p_{12}}} e^{iA_{v_av_b \to p_{23}}} e^{iA_{v_av_b\to p_{30}}}
=
\frac{
\det\left[ \begin{array}{cc}
     \gamma_{\theta_0,\theta_1}(v_b)^\dagger \gamma_{\theta_0,\theta_1}(v_a)&\gamma_{\theta_0,\theta_1}(v_b)^\dagger \gamma_{\theta_1,\theta_2}(v_a)\\
     \gamma_{\theta_1,\theta_2}(v_b)^\dagger \gamma_{\theta_0,\theta_1}(v_a)&\gamma_{\theta_1,\theta_2}(v_b)^\dagger \gamma_{\theta_1,\theta_2}(v_a)\\
\end{array} \right]
}{
\det\left[ \gamma_{\theta_0,\theta_1}(v_b)^\dagger \gamma_{\theta_0,\theta_1}(v_a) \right]
\det\left[ \gamma_{\theta_1,\theta_2}(v_b)^\dagger \gamma_{\theta_1,\theta_2}(v_a) \right]}. 
\end{align}
This is not a real value when the off-diagonal elements $\gamma_{\theta_0,\theta_1}(v_b)^\dagger \gamma_{\theta_1,\theta_2}(v_a)$ and $\gamma_{\theta_1,\theta_2}(v_b)^\dagger \gamma_{\theta_0,\theta_1}(v_a)$ are nonzero.

To eliminate such off-diagonal elements, we proceed as follows. Let $\theta_0^\uparrow \leq \theta_1^\uparrow \leq \theta_2^\uparrow \leq \theta_3^\uparrow$ be the increasing rearrangement of four angles $\theta_0, \theta_1, \theta_2, \theta_3$, and define 
\begin{align}
    &e^{iA^\uparrow_{v_av_b}(j,j+1)} = \det\left[ \gamma_{\theta_j^\uparrow,\theta_{j+1}^\uparrow}(v_b)^\dagger \gamma_{\theta_j^\uparrow,\theta_{j+1}^\uparrow}(v_a) \right],\quad j=0,1,2.
\end{align}
Then, redefine the $U(1)$ connection $A_{v_av_b\to p_{k,k+1}}$ as the diagonal sum of the block $U(1)$ connections $A^\uparrow_{v_av_b}(j,j+1)$s between $\theta_k$ and $\theta_{k+1}$ as in 
\begin{align}
    e^{i\tilde A_{v_av_b\to p_{k,k+1}}}
    := \left[\prod_{j \in \{0,1,2\}: \min(\theta_k,\theta_{k+1})\leq \theta_j^\uparrow \leq \theta_{j+1}^\uparrow \leq \max(\theta_k,\theta_{k+1})} e^{iA^\uparrow_{v_av_b}(j,j+1)}\right]^{\sgn(\theta_k-\theta_{k+1})}
\end{align}
for $k=0,1,2,3$. 
(Note that $k+1 = 0$ when $k+1 = 4$ is interpreted cyclically.)
With this modified $U(1)$ connection $e^{i\tilde A_{v_av_b\to p}}$ contributing to the plaquette $p$ from the edge $v_a v_b$, we define the modified flux over the plaquette $p$ as 
\begin{align}
    \tilde \Phi_p := \mathrm{Arg} \Big[e^{i\tilde A_{v_3v_0 \to p}} e^{i\tilde A_{v_2v_3\to p}} 
    e^{i\tilde A_{v_1v_2\to p}} e^{i\tilde A_{v_0v_1\to p}} \Big].
\end{align}
\end{widetext}
This leads to the refined discretized formula 
\begin{align}
    \tilde W^{\rm dis}_3[g] = \frac{1}{2\pi} \sum_p \tilde \Phi_p. \label{eq:W3_dis}
\end{align}
By construction, it is evident that $e^{2\pi i \tilde W^{\mathrm{dis}}_3[g]} = 1$, which ensures that $\tilde W^{\rm dis}_3[g]$ is an integer.

{\it Model calculation---}
Our formulation extends to computing the winding number for maps $g: X \to GL_N(\mathbb{C})$, where the target space consists of invertible matrices. 
Invertible matrices that cannot be diagonalized, known as "exceptional points," constitute a ring in the three-dimensional parameter space and are stable under minor perturbations. 
To circumvent these exceptional points, one can employ the singular value decomposition $g = U\Sigma V^\dagger$ to derive the unitary matrix $UV^\dagger$ at each vertex within the discretized parameter space. 
We verified our formula (\ref{eq:W3_dis}) with the model $g_0(k_x,k_y,k_z) = t (\sin k_x \sigma_x + \sin k_y \sigma_y + \sin k_z \sigma_z) - i (m+\cos k_x+\cos k_y+\cos k_z)\mathbf{1}_2$ on the three-torus $(k_x,k_y,k_z) \in [-\pi,\pi]^{\times 3}$, employing a cubic lattice with a $20\times 20 \times 20$ mesh. Here, $\sigma_{\mu} \in \{\sigma_x,\sigma_y,\sigma_z\}$ denotes the Pauli matrices. 
We confirmed that the winding number $\tilde W^{\rm dis}_3[g_0]$ equals $-2 \mathrm{sgn}(t)$ for $|m|<1$, $\mathrm{sgn}(t)$ for $1<|m|<3$, and $0$ for $|m|>3$, which is consistent with the direct calculation of the analytic form $W_3[g_0]$. 
Furthermore, we verified that in the model given by $g_1 = (g_0 \oplus e^{\frac{\pi i}{2}} g_0 \oplus u) + v$, where $g_0$ is as defined above, $u$ is a $2 \times 2$ constant random unitary matrix, and $v$ is a $6 \times 6$ constant perturbation, the lattice formula using $\Phi_p$ is not quantized, whereas the formula using $\tilde \Phi_p$ is quantized.

\begin{figure}
    \centering
    \includegraphics[width=\linewidth]{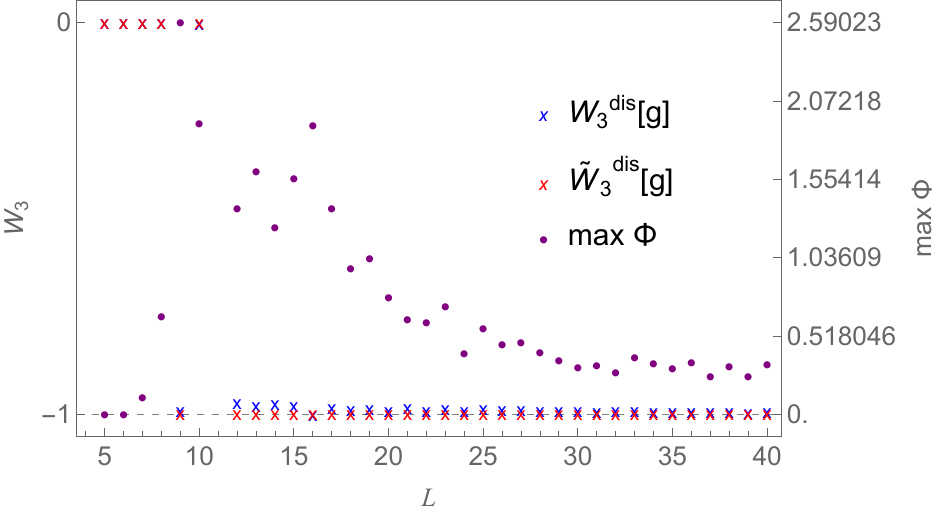}
\caption{
Convergence of the winding number $W_3$ as a function of the lattice size $N$ for a random hopping matrix. 
Blue and red crosses show $W_3^{\mathrm{dis}}[g]$ and $\tilde{W}_3^{\mathrm{dis}}[g]$ computed via $\Phi$ and $\tilde{\Phi}$, respectively. 
The orange circles indicate the maximum plaquette flux $\max_p \Phi_p$. 
The data point at $L=11$ is omitted because there exists a plaquette for which the number of eigenstates within the angular window $(\theta_p^-, \theta_p^+)$ is inconsistent across its vertices.
}
\label{fig:plot}
\end{figure}

We also tested our formulation on randomly generated models. 
Consider a map $g: \mathbb{T}^3 \to \mathrm{Mat}_{2n}(\mathbb{C})$ constructed as a finite-range random hopping. 
Explicitly, we set
\begin{align}
    g(\bm{k}) 
    = \sum_{\bm{r} \in \mathbb{Z}^{\times 3}, |\bm{r}|\leq n_{\rm max}} c_{\bm{r}} e^{-i\bm{k}\cdot\bm{r}} e^{-|\bm{r}|/\xi}
\end{align}
with $c_{\bm{r}} \in \mathrm{Mat}_{2n}(\mathbb{C})$ being independent random matrices, $n_{\rm max}$ the hopping range, and $\xi$ the localization length. 
We may define a map $\tilde{g}: \mathbb{T}^3 \to U(2n)$ as $\tilde{g}(\bm{k}) = U(\bm{k})V(\bm{k})^\dagger$ from the point-wise singular value decomposition $g(\bm{k})=U(\bm{k})\Sigma(\bm{k}) V(\bm{k})^\dagger$.   
However, a randomly generated $g(\bm{k})$ may have stable vortex loop defects where $\det g(\bm{k}) = 0$, since the first homotopy group is nontrivial $\pi_1[GL_{2n}(\mathbb{C})]=\mathbb{Z}$, which obstructs the invertibility of $g(\bm{k})$ over the whole $\mathbb{T}^3$. 
To avoid such defects, we additionally impose the following symmetry constraint 
\begin{align}
    J g(\bm{k})^* J = g(\bm{k}), \quad
J=\mathbf 1_n\otimes\sigma_y .
\end{align}
To do so, we symmetrize $g(\bm{k})$ as
\begin{align}
g(\bm{k})\leftarrow 
\frac12\left(g(\bm{k})+Jg(\bm{k})^*J^{-1}\right).
\end{align}
Then, the unitary-normalized matrix $\tilde{g}(\bm{k})$ belongs to the symplectic group $Sp(n) = \{u \in U(2n) \mid u J u^\top = J\}$, whose low-dimensional homotopy groups vanish, $\pi_k[Sp(n)]=0$ for $k=0,1,2$. 
Therefore, the randomly generated $g(\bm{k})$ is invertible almost surely and possesses a well-defined three-dimensional winding number $W_3$. 
We evaluated $W_3^{\mathrm{dis}}[\tilde{g}]$ and the refined invariant $\tilde W_3^{\mathrm{dis}}[\tilde{g}]$ on an $L\times L\times L$ cubic lattice of $\mathbb{T}^3$ while varying the grid length $N$ for a randomly chosen nearest-neighbor hopping model ($n_{\rm max}=1$) whose three-dimensional winding number is $W_3 = -1$. 
As shown in Fig.~\ref{fig:plot}, the results confirm that while the unmodified formulation using $\Phi_p$ generally fails to yield a quantized value, the modified formula using $\tilde{\Phi}_p$ consistently produces an integer.
Both converge to $W_3 = -1$ as $L$ increases. 
The maximum plaquette flux $\max_p \Phi_p$ decreases toward zero, demonstrating the improvement of the lattice approximation with increasing $L$. 
A sample Mathematica code and the hopping parameters used for Fig.~\ref{fig:plot} are available as ancillary files on arXiv.

{\it Summary and discussion---}
In this work, we presented a formulation for calculating the three-dimensional winding number $W_3[g]$ for smooth maps $g: X \to U(N)$, utilizing a discrete approximation of the manifold $X$. 
Our approach allows for the computation of $W_3[g]$ exclusively through the diagonalization of matrices $g(v)$ at a finite number of vertices. 
Compared to the preceding algorithm~\cite{Hockendorf_2017}, our formulation using $\theta$-gaps circumvents the need to track and match individual eigenvectors between adjacent sites. This simplifies the numerical implementation and allows for direct application to systems with accidental or symmetry-enforced degeneracies. 

Regarding the practical implementation, it should be noted that while the unmodified flux $\Phi_p$ is determined solely by the information on the plaquette itself, the modified flux $\tilde{\Phi}_p$ requires the $\theta$-gap information from the four cubes adjacent to each edge to strictly ensure integer quantization. 
However, our numerical tests indicate that the sum of the unmodified flux $\Phi_p$ is almost always quantized for sufficiently fine grids, making it a sufficient and more straightforward choice for practical numerical evaluations. 

Discrete formulations that explicitly compute topological invariants are currently limited in scope. Examples such as instanton numbers represented by the second Chern numbers, higher-dimensional winding numbers, and degrees of maps to more general symmetric spaces have yet to be explored. 
We look forward to future studies shedding more light on these topics.

\begin{acknowledgments}
I thank Tsuneya Yoshida for pointing out the error in the v1 version regarding the quantization of the lattice formula for the three-dimensional winding number.
This work was supported by JST CREST Grant Number JPMJCR19T2, and JSPS KAKENHI Grant Numbers JP22H05118 and JP23H01097.
\end{acknowledgments}

\bibliography{ref}

\end{document}